\documentclass{eptcs}

\usepackage{breakurl}

\usepackage{amsmath}
\usepackage{pslatex}
\usepackage{amssymb}
\usepackage{amsthm}

\usepackage{stmaryrd}
\usepackage{graphicx}
\usepackage{marvosym}
\usepackage{color}
\usepackage{pdfsync}
\usepackage{wasysym}

\newtheorem{theorem}{Theorem}[section]
\newtheorem{lemma}[theorem]{Lemma}
\newtheorem{definition}[theorem]{Definition}
\newtheorem{example}{Example}[section]
\newtheorem{remark}{Remark}[section]

\graphicspath{{figures/}} \DeclareGraphicsExtensions{.pdf,.jpg}
\usepackage[center]{subfigure}

\newif\ifemi
  \emifalse
  \emitrue

\newif\ifarticle
  \articletrue
  \articlefalse

\newif\iftesi
  \tesitrue
  \tesifalse

\newif\ifnat
  \nattrue
  \natfalse

\newif\ifpar
  \partrue
  \parfalse

\newif\ifllncs
  \llncsfalse
  \llncstrue

\newif\ifdraft
  \draftfalse
  \drafttrue

\newif\ifthm
  \thmfalse
  \thmtrue

\newif\ifqapl
  \qaplfalse

\newif\ifmod
  \modtrue
  \modfalse

\newif\ifentcs
  \entcstrue
  \entcsfalse

\newcommand{\emptysys}{\mathbf 0}
\newcommand{\rname}[1]{(\mbox{\sc #1})}

\def\act#1#2{#1 @ #2}

\newcommand{\Real}[1]{\mathrm{Real}}

\newcommand{\coco}{\mbox{\ensuremath{\mathrm{CO}_2}\hspace{2pt}}}

\newcommand{\pmv}[1]{\ensuremath{\mathsf{#1}}}
\newcommand{\tuple}[1]{\ensuremath{\langle {#1} \rangle}}

\newcommand{\names}{\ensuremath{\mathcal{N}}}
\newcommand{\snames}{\names_S}
\newcommand{\pnames}{\names_P}

\newcommand{\vars}{\mathcal V}
\newcommand{\svars}{\vars_S}
\newcommand{\pvars}{\vars_P}

\newcommand{\atom}[1]{\textup{\textsf{#1}}}

\newcommand{\fact}[2]{\mathsf{do}_{#1}\,{#2}}
\newcommand{\tell}[2]{\mathsf{tell}_{#1}\,{#2}}
\newcommand{\sys}[2]{{#1} [{#2}] }

\newcommand{\freeze}[2]{\downarrow_{#1}{#2}}

\newcommand{\coimp}{\twoheadrightarrow}

\newcommand{\honest}[2]{{#1} \smiley {#2}}

\newcommand{\sep}{\ \bnfmid\ }

\newcommand{\redrule}[2]{
\prooftree
{#1}
\justifies
{#2}
\endprooftree
}
\newcommand{\thaw}{\uparrow}
\newcommand{\agreement}[4]{{#1} \vartriangleright_{#3}^{#4} {#2}}

\newcommand{\dmid}{\mid\hspace{-0pt}\mid}

\newcommand{\compile}[2]{\ifthenelse{\equal{#1}{yes}}{#2}{}}

\newcommand{\cf}[2]{
  \fontsize{#1}{#1}{\selectfont{#2}}
}
\ifemi
\usepackage{ulem}
\newcommand{\emi}[1]{{\marginpar{\cf{6}{{#1}}}}}
\newcommand{\emic}[2]{\par
  \definecolor{shadecolor}{rgb}{1,0.99,0.9}
  \fcolorbox{red}{shadecolor}{\parbox{\linewidth}{ 
      \begin{description}
      \item[{\color{blue} #2}]{\sf #1}
      \end{description}}}
}

\else
\newcommand{\emi}[1]{}
\newcommand{\emic}[2]{}

\fi

\ifemi

\else

\fi

\ifmod

\else

\fi

\ifqapl
\newcommand{\dom}[1]{\mathrm{dom}({#1})}

\else
\newcommand{\dom}[1]{\operatorname{dom} {#1}}

\fi

\newcommand{\proofend}{\mbox{$\Box$}}

\newcommand{\mmdef}{\mbox{$\;\stackrel{\textrm{\tiny def}}{=}\;$}}

\ifpar
\else
\newcommand{\defeq}{\mathrel{\mathop{=}\limits^{\rm def}}}
\fi
\newcommand{\nil}{\mathbf{0}}
\newcommand{\zero}{\mathbf{0}}

\newcommand{\fn}[1]{\mathrm{fn}(#1)}

\newcommand{\fv}[1]{\mathrm{fv}(#1)}

\ifpar

\else

\fi

\ifarticle
\newtheorem{theorem}{Theorem}[section]
\newtheorem{definition}{Definition}[section] 
\newtheorem{proposition}{Proposition}[section] 
\newtheorem{lemma}{Lemma}[section]
\newtheorem{corollary}{Corollary}[lemma] 
\newtheorem{remark}{Remark}[section]
\newtheorem{observation}{Observation}[section]
\newtheorem{notation}{Notation}[section]
\newtheorem{example}{Example}[section]
\else
\ifthm
\else
\newtheorem{theorem}{Theorem}[section]
\newtheorem{definition}{Definition}[section]

\newtheorem{example}{Example}[section]

\fi
\fi

\def \overunderstackrel#1#2#3{\mathrel{\mathop{#1}\limits^{#2}_{#3}}}
\def \overstackrel#1#2{\mathrel{\mathop{#1}\limits^{#2}}}

\newcommand{\subs}[2]{\{^{#1}/_{#2}\}}

\newcommand{\bnfmid}{\;\big|\;}

\newcommand{\imp}{\rightarrow}

\newcommand{\ask}[2]{\mathsf{ask}_{{#1}}\,{#2}}
\newcommand{\fuse}[2]{\mathsf{fuse}_{{#1}}\,{#2}}
 \newcommand{\says}{\ensuremath{\;\mathit{says}\;}}

\newcommand{\pcl}{\textup{PCL\;}}

\newcommand{\pclminus}{\ensuremath{\,\pcl^{\!\!\!-}}}

\newcommand{\setenum}[1]{\{#1\}}
\newcommand{\setcomp}[2]{\{{#1} \;\mid\; {#2}\}}
\newcommand{\entails}{\vdash}

\input prooftree

\newcommand{\mytitle}[0]{Contracts in distributed systems}

\title{\mytitle}

\author{Massimo Bartoletti\inst{1} \and \inst{2} \and \inst{3}}

\author{Massimo Bartoletti
\institute{Dipartimento di Matematica e Informatica, Universit\`a degli Studi di Cagliari, Italy 
}
\and
Emilio Tuosto
\institute{Department of Computer Science,
University of Leicester, UK 
}
\and
Roberto Zunino
\institute{DISI-Universit\`a degli Studi di Trento and COSBI, Italy
}
}

\def\titlerunning{\mytitle}
\def\authorrunning{M. Bartoletti, E. Tuosto, R. Zunino}

\begin{document}

\maketitle
\begin{abstract}
  We present a parametric calculus for contract-based computing in
  distributed systems.  By abstracting from the actual contract
  language, our calculus generalises both the
  \emph{contracts-as-processes} and \emph{contracts-as-formulae}
  paradigms.  The calculus
  features primitives for advertising contracts, for reaching
  agreements, and for querying the fulfilment of contracts.
  Coordination among principals happens via multi-party sessions,
  which are created once agreements are reached.  We present two
  instances of our calculus, by modelling contracts as ($i$) processes
  in a variant of CCS, and ($ii$) as formulae in a logic.  With the
  help of a few examples, we discuss the primitives of our calculus,
  as well as some possible variants.
\end{abstract}

\newcommand{\cem}[2]{#2}

\section{Introduction}
What are contracts for distributed services? How should they be used?
These questions are intriguing not only researchers but also
practitioners and vendors.
In fact, contracts are paramount
for correctly designing, implementing, and composing distributed
software services.
In such settings, contracts are used at different levels of abstraction,
and with different purposes.
Contracts are used to model the possible interaction patterns of
services, with the typical goal of composing those services only which
guarantee deadlock-free interactions.  At a different level of
abstraction, contracts are used to model Service Level Agreements
(SLAs), specifying what has to be expected from a service, and what
from the client.  Also in this case, a typical goal is that of
matching clients and services, so that they agree on the
respective rights and obligations.

Contracts have been investigated from a variety of perspectives
and using a variety of different formalisms and analysis techniques,
ranging from c-semirings~\cite{Buscemi07transactional,Buscemi07ccpi,Ferrari06logic}, to
behavioural types~\cite{Bravetti07sc,Carpineti06basic,Castagna09contracts},
to formulae in suitable logics~\cite{Artikis09jlap,BZ10lics,PrisacariuS07formal}, 
to categories~\cite{Cardone11geometry}, \emph{etc}.
This heterogeneous ecosystem of formalisms makes it difficult to understand
the essence of those methods, and  how they are related.

As a first step towards remedying this situation, we propose a generic
calculus for Contract-Oriented COmputing (in short, \coco\!).
By abstracting away from the actual contract language, our calculus
can encompass a variety of different contract paradigms.
We provide a common set of primitives for computing with contracts:
they allow for advertising and querying contracts, for reaching
agreements, and for fulfilling them with the needed actions.
All these primitives are independent from the chosen
language of contracts, and they only pivot on some general
requirements fixed in the \emph{contract model} proposed here.

A remarkable feature of our approach is that contracts are not
supposed to be always respected after they have been stipulated.  
Indeed, we can model the quite
realistic situation where promises may be possibly reneged.  
Therefore, in \coco contracts are not discharged after they have been
used to couple services and put them in a session, as usually done e.g.\ 
in the approaches dealing with behavioural types. In our approach,
contracts are also used to drive computations after sessions have been
established, e.g.\ to detect violations and to provide the agreed
compensations.

\paragraph{Synopsis.}

The overall contribution of the paper is a calculus 
for computing with contracts in distributed systems.
The calculus is designed around two main principles.

The first is the separation of concerns between the way contracts are
modelled and the way they are used in distributed computations.
Indeed, we abstract from the actual contract language by only imposing
a few general requirements.
In this way, we envisage our calculus as a generic framework which can
be tuned by instantiating the contract model to concrete
formalisations of contracts.
In \S~\ref{sec:contract-model} we present the abstract contract model,
followed by two concretisations:
in \S~\ref{sect:contracts-as-processes} we adopt the
contracts-as-processes paradigm whereby CCS-like processes
represent contracts that drive the behaviour of distributed
participants;
in \S~\ref{sect:contracts-as-formulae} we embrace instead the
contracts-as-formulae paradigm, by instantiating our calculus
with contracts expressed in a suitable logic.
We relate the two concrete models in~\S~\ref{sec:comparison}, 
first with the help of a few examples, and then by showing that
contracts-as-formulae, expressed in a significant fragment of our logic, 
can be suitably encoded into contracts-as-processes (Theorem~\ref{th:pcl-ccs}).

The second design principle of our calculus
is that its primitives must be reasonably implementable in a
distributed setting.
To this purpose, we blend in~\S~\ref{sec:co2} a few primitives inspired 
by Concurrent Constraint Programming (CCP~\cite{Saraswat91cc}) 
to other primitives inspired by session
types~\cite{honda.vasconcelos.kubo:language-primitives}.
The key notions around which our primitives are conceived are
\emph{principals} and \emph{sessions}.
The former represent distributed units of computation that can
advertise contracts, execute the corresponding operations, and
establish/check agreements.
Each agreement corresponds to a fresh session, containing
rights and obligations of each stipulating party.
Principals use sessions to coordinate with each other and fulfil their
obligations.
Also, sessions enable us to formulate a general notion of
``misbehaviour'' which paves the way for automatic verification.
We finally suggest possible variants of our primitives and of the contract model
(\S~\ref{sec:variants}).

\paragraph{Related Work.} \label{sec:related-work}

Multi-party session types~\cite{mps} are integrated in~\cite{bhty10}
with decidable fragments of first-order logic (e.g., Presburger
arithmetic) to transfer the design-by-contract of object-oriented
programming to the design of distributed interactions.
\cem{In our approach we}{We} follow a methodologically opposite direction.
In fact, in~\cite{bhty10} one starts from a \emph{global assertion}
(i.e., global choreography and contracts) to arrive
to a set of \emph{local assertions}; \cem{}{distributed} processes
abiding with local assertions are guaranteed to have correct
\cem{distributed}{} interactions (and monitors can be synthesized from local
assertions to control execution in untrusted settings).
In our framework instead, 
a principal declares its contract
independently of the others and then advertises it; 
a \coco
primitive tries then to harmonise contracts by searching for a suitable
agreements.
In other words, one could think of our approach as based on
orchestration rather than choreography.
%
The same considerations above apply to~\cite{mcneile10} where protocol
modelling (state machines with memory) \cem{are used to
  represent}{represents} global choreographies.
There, contracts are represented as \cem{state machines that run in
  parallel}{parallel state machines} (according to a CSP-like
semantics).
Basically, the contract model of~\cite{mcneile10} coincides with its
choreography model.

In {\sf cc}-pi~\cite{Buscemi07ccpi}, CCP is mixed with communication
through name fusion. In this model, \cem{SLA is obtained by merging
  the constraints of the involved parties, which represent their
  requirements.}{involved parties establish SLA by merging the
  constraints representing their requirements.}  Constraints are
values in a c-semiring advertised in a global store.  It is not
permitted to merge constraints making the global store inconsistent,
since an agreement cannot be reached in that case.  
\cem{By comparison}{Conversely,
\coco envisages} contracts as binding promises rather than
requirements.  
Actually, even if a principal \pmv A tells an absurdum, this
will result in a contract like: ``$\pmv A$ is stating a contradiction'' 
added to the environment. 
When this happens, our approach is not ``contracts are
inconsistent, do not open a session'', but rather ``\pmv A is
promising the impossible, she will not be able to keep her promise,
and she will be blamed for that''.
%
The {\sf cc}-pi calculus is further developed
in~\cite{Buscemi07transactional} to include long running
transactions and compensations.  There, besides the global constraint
store, \cem{the calculus features}{} a local store for each transaction \cem{}{is featured}. Local
inconsistencies are then used to trigger compensations.
In \coco, compensations do not represent exceptional behaviour to be
automatically triggered by inconsistencies; rather, compensations fall
within ``normal'' behaviour and have to be spelt out inside contracts.
Indeed, after a session has been established, each honest principal
\pmv A either maintains her promises, or she is culpable of a violation;
she cannot simply try to execute arbitrary compensations in place of
the due actions.  Of course, other principals may deem this promise
too weak and avoid establishing a session with \pmv A.

In \cite{Coppo08structured} a calculus is proposed to model SLAs which
combines $\pi$-calculus communication, concurrent constraints, and
sessions.  There, the constraint store is global and sessions are
established between two processes whenever the stated requirements are
consistent. Interaction in sessions happens through communication and
label branching/selection. A type system is provided to guarantee safe
communication, although not ensuring progress.  Essentially, the main
role of constraints in this calculus is that of driving session
establishment. Instead, in \coco the contracts of an agreement leading
to a session are still relevant e.g.~to detect violations.

A ``boolean'' notion of compliance between two contracts is
introduced in~\cite{Castagna09mobile}: either the contract of the
client and one of the service are compliant, or they are not. In
Ex.~\ref{ex:compliance} we discuss a ``multi-level'' notion of
compliance encompassing more than two contracts. Also,
in~\cite{Castagna09mobile} not compliant contracts, may become
compliant by adjusting the order of asynchronous actions. When this is
possible, an orchestrator can be synthesised from the client and
service contracts. In some sense, the orchestrator acts as an
``adapter'' between the client and the service. In our approach, the
orchestrator behaves as a ``planner'' which finds a suitable set of
contracts and puts in a session all the principals involved in these
contracts. 

\coco takes inspiration from~\cite{BZ10lics,BZ10ice}.  There,
the contract language is the logic \pcl\!, and contracts are recorded into a
global constraint store.  \coco instead features local environments
for principals and sessions to enable possible distributed
implementations.

Our approach differs from those discussed above,
as well as from all the other approaches we are aware of 
(e.g.~\cite{Bravetti07sc,Carpineti06basic,Castagna09contracts}), 
w.r.t.~two general principles.
First, we depart from the common principle that contracts are always respected
after \cem{they are stipulated}{their stipulation}. 
\cem{In our contract model we}{We} represent instead the more realistic situation
where promises are not always maintained.
As a consequence, in \coco we do not discard contracts
after they have been used to couple services and put them in a session,
as done e.g.\ in all the approaches dealing with behavioural types.
In our approach, contracts are also used to drive computations after 
sessions have been established (cf.\S~\ref{sec:co2}), 
e.g.\ to detect violations and to provide the agreed compensations.

The second general difference is that \coco smoothly
allows for handling contracts-as-processes 
(cf.\S~\ref{sect:contracts-as-processes}). 
To the best of our knowledge,
it seems hard to accomodate these contracts within frameworks
based on constraint systems (\cite{BZ10ice,Coppo08structured}), 
logics (\cite{BZ10lics,bhty10}),
or \mbox{c-semirings} (e.g.~\cite{Buscemi07ccpi,Buscemi07transactional,Ferrari06logic}).


\section{An abstract contract model} \label{sec:contract-model}

We now sketch the basic ingredients of a generic contract model, before
providing a formal definition.

We start by introducing some preliminary notions and definitions; 
some of them will only be used later on in \S~\ref{sec:co2}.
\emph{Principals} are those agents which may
advertise contracts, establish agreements, and realise them.
\emph{Sessions} are created upon reaching an agreement, and provide
the context in which principals can interact to fulfil their contracts.
Let $\names$ and $\vars$ be countably infinite, disjoint sets
of names and variables, respectively.
Assume $\names$ partitioned into two infinite sets
$\pnames$ 
and $\snames$,
for names of principals and of sessions, respectively.
Similarly, $\vars$ is partitioned into infinite sets $\pvars$
and $\svars$ 
for variable identifiers of principals and sessions.
A substitution is a partial map $\sigma$ from $\vars$ to $\names$; we
write $u \in \dom \sigma$ when $\sigma$ is defined at $u$, and require
that $\sigma$ maps $a \in \dom \sigma \cap \pvars$ to $\pnames$ and $s
\in \dom \sigma \cap \svars$ to $\snames$.

Our main notational conventions are displayed in
Table~\ref{tbl:notation}.

\begin{table}[t]
  \hrulefill
  \small
  \[\begin{array}{c@{\hspace{20pt}}c}
    \begin{array}{ll}
      n, m, \ldots \in \names \hspace{30pt} & \text{names, union of:} \\
      \hspace{12pt} \pmv A, \pmv B,\ldots \in \pnames & \hspace{12pt} \text{principal names} \\
      \hspace{12pt} s,t,\ldots \in \snames & \hspace{12pt} \text{session names} \\
      \vars & \text{variables, union of:} \\
      \hspace{12pt} a,b,\ldots \in \pvars & \hspace{12pt} \text{principal variables} \\
      \hspace{12pt} x,y,\ldots \in \svars & \hspace{12pt} \text{session variables} \\
      u, v, \ldots \in \names \cup \vars & \text{names or variables}
    \end{array}
    &
    \begin{array}{ll}
      A, B, \ldots \in \pnames \cup \pvars & \text{principal names/variables} \\
     \atom{a},\atom{b},\ldots \in \mathcal{A} \hspace{0pt} & \text{atoms} \\
     \tuple{A_1 \says \atom{a}_1,\ldots,A_j \says \atom{a}_j} \hspace{12pt} & \text{action} \\
      c,c',\ldots \in \mathcal{C} & \text{contracts} \\
      C & \text{multisets of contracts} \\
      \phi, \psi, \ldots \in \Phi & \text{observables} \\
      \mu, \mu', \ldots & \text{labels}
    \end{array}
  \end{array}
  \]
  \hrulefill
  \caption{Notation \label{tbl:notation}}
\end{table}

The first ingredient of our contract model is 
a set $\mathcal{C}$ of \emph{contracts}.
We are quite liberal about it: we only require that 
$A \says c \in \mathcal{C}$ for all principals $A$
and for all $c \in \mathcal{C}$.
The contract $A \says c$ can be thought of as 
``$c$ is advertised by $A$''.
A labelled transition relation $\xrightarrow{\mu}$ on contracts models
their evolution under the actions performed by principals.

Two further ingredients are a set $\Phi$ of \emph{observables}
(properties of contracts) and an entailment relation $\vdash$ between
contracts and observables.
Note that we keep distinct contracts from observables in our framework. 
This has the same motivations as the traditional distinction between behaviours
(systems) and their properties (formulae predicating on behaviours), 
which brought in plenty of advantages in the design/implementation of systems.

The last ingredient of our contract model is a relation $\honest{}{}$
between contracts and principals.
We write $\honest{C}{\pmv{A}}$ to mean that, with respect to contracts
$C$, all the obligations of the principal $\pmv A$ have been fulfilled.

Def.~\ref{def:contract-model} formalises the above concepts.

\begin{definition} \label{def:contract-model}
A \emph{contract model} is a tuple
  $\tuple{\mathcal{C}, \mathcal{A}, \xrightarrow{}, \Phi, \; \vdash,\honest{}{}}$ where
  \begin{itemize}
  \item $\mathcal{C}$ is a \emph{set of contracts}, 
    forming a subalgebra of \cem{the}{a} term-algebra $T_{\vars\cup\names}(\Sigma)$
    for some signature $\Sigma$ which includes the operations 
    $u \says \_\,$ for each $u \in \pvars \cup \pnames$
  \item $\mathcal{A}$ is a set of \emph{atoms} 
    (ranged over by $\atom{a}, \atom{b}, \ldots$)
  \item $C \xrightarrow{\mu} C'$ is a labelled transition relation
    over finite multisets on $\mathcal{C}$. 
    The set of labels comprises {\em actions}, i.e.\ 
    tuples of the form 
    \tuple{A_1 \says \atom{a}_1,\ldots, A_j \says \atom{a}_j}
  \item $\Phi$ is a set of \emph{observables}, forming a subalgebra of
    \cem{the}{a} term-algebra $T_{\vars \cup \names}(\Sigma')$ for
    some signature $\Sigma'$
  \item $\vdash$ is a \emph{contract entailment} relation between finite 
    multisets of $\mathcal{C}$ and $\Phi$
  \item $\honest{}{}$ is a \emph{contract fulfilment} relation between 
    finite multisets of $\mathcal{C}$ and principals.
  \end{itemize}
\end{definition}

\begin{example} \label{ex:sale:abstract}
We illustrate the contract model with the help of an informal example.
A seller $\pmv{A}$ and a buyer $\pmv{B}$
stipulate a contract $c_0$, which binds
$\pmv{A}$ to ship an item after $\pmv{B}$ has paid.
Let $\atom{pay}$ be the atom which models the action of paying.
The transition $c_0 \xrightarrow{{\pmv B} \says \atom{pay}} c_1$ 
models the evolution of $c_0$ into a contract $c_1$ where
$\pmv A$ is obliged to \cem{pay}{ship}, while $\pmv B$ has no more
duties.
Now, let $\phi$ be the observable ``$\pmv A$ must ship''.  
Then, we would have $c_0 \not\vdash \phi$, 
because $\pmv A$ does not have to ship anything yet, 
while $c_1 \vdash \phi$, because $\pmv B$ has paid and so $\pmv A$ must ship.
It would not be the case that $\honest{c_1}{\pmv A}$,
because \pmv B has paid, while
\pmv A has not yet fulfilled her obligation to ship.
\end{example}

We remark that the use of term-algebras in
Def.~\ref{def:contract-model} allows us to smoothly apply variable
substitutions 
to contracts and observables.
 Accordingly, we assume defined the sets $\fv c$ and $\fv \phi$ of (free) variables of contracts and observables.
Note that actions 
are not required to be in $\mathcal{C}$.
Depending on the actual instantiation of the contract model, it can
be useful to include them in $\mathcal{C}$, so that contracts
can record the history of the past actions.

\subsection{Contracts as processes}\label{sect:contracts-as-processes}

The first instance of our contract model appeals to the
contracts-as-processes paradigm.
A contract is represented as a CCS-like
process~\cite{Milner89ccs}, the execution of which dictates obligations to
principals.

\begin{definition} \label{def:ccs-model}
  We define a \emph{contracts-as-processes} language as follows
  \begin{itemize}
  \item $\mathcal{C}$ is the set of process terms defined by 
    the following grammar:
    \[
    c ::= \textstyle \sum_i \atom{a}_i.c_i \ \sep \ A \says c \ \sep\ c \mid c\ \sep \ X \\
    \]
    and $\Sigma$ is the signature corresponding to the syntax above;
    in this section, multisets of contracts are identified with their
    parallel composition, 
    and accordingly we use the metavariable $c$ to denote them.
    We assume variables $X$ to be defined through (prefix-guarded recursive) equations.
  \item $\mathcal A$ is the union of three disjoint
    sets: the ``inputs'' (ranged over by $\atom{a}^-$), the
    ``outputs'' (ranged over by $\atom{a}^+$), and the ``autonomous
    activities'' (ranged over by $\atom{a}^0$).
  \item $\xrightarrow{\mu}$ is the least relation closed under the rules in
    Table~\ref{tbl:ccs-contracts} and structural equivalence $\equiv$
      (defined with the usual rules and $A \says \nil \equiv
      \nil$, where $\nil$ denotes the empty sum and trailing occurrences
      of $\nil$ may be omitted).
  \item $\Phi$ is the set of LTL~\cite{Emerson90temporal} formulae
    (on a signature $\Sigma'$), where the constants are the atoms in $\mathcal{A}$. 
  \item $c \vdash \phi$ (for closed $c$ and $\phi$) holds when $c
    \models_{LTL} \phi$ according to the standard LTL semantics
    where\cem{the semantics of prime formulae}{, given a generic trace
      $\eta$ of $c$, the semantics of atoms is:}
    \[
    \begin{array}{rcl}
      \eta \models \atom{a}^0 & \iff & \exists A,\eta'.\; \eta=\tuple{A \says \atom{a}^0}
      \ \eta'
      \\[2pt]
      \eta \models \atom{a}^+ \iff 
      \eta \models \atom{a}^- & \iff &
      \exists A_1,A_2,\eta'.\; \eta=\tuple{A_1 \says \atom{a}^-,A_2 \says \atom{a}^+}
      \ \eta'
    \end{array}
    \]
  \item $\honest{c}{A}$ holds iff for all $c', c''$ such that $c \equiv
    (A \says c') \mid c''$ we have that $c' \equiv \nil$.
  \end{itemize}
\end{definition}

\begin{table}[!t]
  \hrulefill
  \small
  \[\hspace{-5pt}\begin{array}{c}
    \mu ::= \atom{a} \sep \tuple{A \says \atom{a}^0} \sep 
    \tuple{A \says \atom{a}^-,A \says \atom{a}^+}
    \\[1pc]
    \sum_i \atom{a}_i.c_i \ \xrightarrow{\atom{a}_i} c_i
    \hspace{10pt}\rname{\sf Sum}
    \qquad 
    \redrule
    {c_1 \xrightarrow{\mu} c_1'}
    {c_1 \mid c_2 \xrightarrow{\mu} c_1' \mid c_2}
    \hspace{10pt}\rname{\sf Par}
    \qquad 
    \redrule
    {X \mmdef c \qquad c \xrightarrow{\mu}c' }
    {X \xrightarrow{\mu} c'}
    \hspace{10pt}\rname{\sf Def}
    \\[20pt]
    \redrule
    {c \xrightarrow{\atom{a}^0} c' \quad
     \mu = \tuple{A \says \atom{a}^0}
    } 
    {A \says c \xrightarrow{\mu} A \says c'}
    \hspace{5pt}\rname{\sf Auto}
    \qquad 
    \redrule
    {c_1 \xrightarrow{\atom{a}^-} c_1' \quad
     c_2 \xrightarrow{\atom{a}^+} c_2' \quad
     \mu = \tuple{A_1 \says \atom{a}^-,\ A_2 \says \atom{a}^+}
    }%
    {A_1 \says c_1 \mid A_2 \says c_2 
      \xrightarrow{\mu}
      A_1 \says c_1' \mid A_2 \says c_2'}
    \hspace{5pt}\rname{\sf Com}
  \end{array}
  \]
  \hrulefill
  \vspace{-0pt}
  \caption{\label{tbl:ccs-contracts}Labelled transition relation of
    contract-as-processes}
\end{table}

We briefly comment on the rules in Table~\ref{tbl:ccs-contracts}.
Intuitively, the relation $\xrightarrow \mu$ either carries labels of
the form $\tuple{A \says \atom{a}^0}$, which instruct $A$ to fulfil
the obligation $a^0$, or labels $\tuple{A_1 \says \atom{a}^-,\
  A_2 \says \atom{a}^+}$, which require participants $A_1$
and $A_2$ to fulfil the obligations $\atom a^-$ and $\atom a^+$,
respectively.
Rules \rname{Sum}, \rname{Par}, and \rname{Def} are standard.
By rule \rname{Auto}, a contract willing to perform an autonomous
action $\atom a^0$ can do so and exhibit the label $\tuple{A \says \atom{a}^0}$.
Rule \rname{Com} is reminiscent of the synchronisation mechanism of
CCS; when two complementary actions $\atom a^-$ and $\atom a^+$ can be
fired in parallel, then the parallel composition of contracts emits
the tuple $\tuple{A_1 \says \atom{a}^-,\ A_2 \says \atom{a}^+}$.
Note that the rules in Table~\ref{tbl:ccs-contracts} give semantics to
\emph{closed} contracts, i.e.\ contracts with no occurrences of
free variables.

\begin{example}\label{ex:sale:ccs}
Recall the buyer-seller scenario from Ex.~\ref{ex:sale:abstract}.
The seller $\pmv A$ promises to ship an item if 
buyer 
$\pmv{B}$ promises to pay.
The buyer $\pmv{B}$ promises to pay.
The contracts of $\pmv{A}$ and $\pmv{B}$ are as follows:
\begin{align*}
c_{\pmv A} & = {\pmv A} \says \atom{pay}^-.\atom{ship}^0
\hspace{40pt} &
c_{\pmv B} & = {\pmv B} \says {\atom{pay}^+}
\end{align*}
A possible computation is then:
\(
  c_A \mid c_B 
\xrightarrow{\tuple{\pmv A \says \atom{pay}^-,\ \pmv B \says \atom{pay}^+}}
{\pmv A} \says \atom{ship}^0 \ |\ \nil
\xrightarrow{\tuple{\pmv A \says \atom{ship}^0}}
\nil
\).
\end{example}

It is evident that the contract of $\pmv B$ in Ex.~\ref{ex:sale:ccs} is
rather naive; the buyer pays without requiring any guarantee to the
seller (cf. Ex.~\ref{ex:sale:2}).
A possible solution is to use a (trusted) escrow service.

\begin{example}\label{ex:escrow}
In the same scenario of Ex.~\ref{ex:sale:ccs}, consider an escrow
service \pmv E which mediates between \pmv A and \pmv B.
Basically, \pmv A and \pmv B trust the escrow service $\pmv E$,
and they promise to ship to \pmv E and to pay \pmv E, respectively.
The escrow service promises to ship to \pmv B and to pay \pmv A only
after both the obligations of \pmv A and \pmv B have been fulfilled.
The contracts of $\pmv A$, $\pmv B$, and
$\pmv E$ are defined as follows:
  \[\begin{array}{rcl}
    c_{\pmv A} & \mmdef & \pmv A \says \atom{shipE}^+.\atom{pay}^-
    \\
    c_{\pmv B} & \mmdef & \pmv B \says \atom{payE}^+.\atom{ship}^-
  \end{array}
  \qquad
  \begin{array}{rcl}
    c_{\pmv E} & \mmdef \;\;\; \pmv E \says \hspace{-8pt}  & \atom{shipE}^-.\ \atom{payE}^-.\ ( \atom{pay}^+ \mid \atom{ship}^+) \; + \\
    &         & 
     \atom{payE}^-.\ \atom{shipE}^-.\ (\atom{pay}^+ \mid  \atom{ship}^+)
  \end{array}
\]
\end{example}

\subsection{Contracts as formulae}\label{sect:contracts-as-formulae}

For the second specialization of our generic model,
we choose the contract logic \pcl\!\!~\cite{BZ10lics}.
A comprehensive presentation of \pcl is beyond the scope of this paper,
so we give here just a brief overview, and we refer the reader 
to~\cite{BZ10lics,PCLtr} for more details.

\pcl extends intuitionistic propositional logic IPC~\cite{Troelstra}
with the connective~$\coimp$, called \emph{contractual implication}.
Differently from IPC,
a contract $\sf b \coimp a$ implies $\sf a$ not only when $\sf b$ is true, 
like IPC implication, but also in the case that a ``compatible'' 
contract, e.g.\ $\sf a \coimp b$, holds.
So, \pcl allows for a sort of ``circular'' assume-guarantee reasoning,
summarized by the theorem 
$\vdash \sf (b \coimp a) \;\land\; (a \coimp b) \;\imp\; a \land b$.
Also, \pcl is equipped with an indexed lax modality
$\_ \says \_$, similarly to
the one in~\cite{Garg08modal}.

The proof system of \pcl extends that of IPC with the following axioms,
while remaining decidable:
\begin{align*}
& \top \coimp \top 
&& \phi \imp (A \says \phi) \\
& (\phi \coimp \phi) \imp \phi
&& (A \says A \says \phi) \imp A \says \phi \\
& (\phi' \imp \phi) \imp (\phi \coimp \psi) \imp (\psi \imp \psi') \imp (\phi' \coimp \psi')
&& (\phi \imp \psi) \imp (A \says \phi) \imp (A \says \psi) 
\end{align*}

Following Def.~\ref{def:contract-model}, we now define a contract language which builds upon \pcl\!.

\vbox{
\begin{definition}
We define a {\em contracts-as-formulae} language as follows:
\begin{itemize}

\item $\mathcal{C}$ is the set of \pcl formulae. 
Accordingly, $\Sigma$ comprises all the atoms
$\mathcal{A}$ (see below), all the connectives of \pcl\!, and the
$\_ \says \_$ modality.

\item $\mathcal{A}$ is partitioned in {\em promises},
  written as $\atom{a}$, and {\em facts}, written as $!\atom{a}$.

\item The labelled relation $\xrightarrow{\mu}$ is defined by the rule:
\(
 C \;\;\;\xrightarrow{\tuple{A \says \atom{a}}}\;\;\; 
  C ,\ A \says \atom{a},\ 
  A \says !\atom{a}
\)

\item $\Phi = \mathcal{C}$, and $\Sigma'=\Sigma$.

\item $\vdash$ is the provability relation of \pcl\!.

\item $\honest{C}{A}$ holds iff
  $C\vdash A \says \atom{a}$ implies $C\vdash A \says
  !\atom{a}$, for all promises $\atom{a}$, i.e.\ each obligation for
  $A$ entailed by $C$ has been fulfilled.

\end{itemize}
\end{definition}}

Note that the definition of $\xrightarrow{\mu}$ allows principals to perform any actions: 
the result is that $C$ is augmented with the corresponding fact $!\atom{a}$.
We include the promise $\atom{a}$ as well, following the intuition
that a fact may safely imply the corresponding promise.

\begin{example} \label{ex:sale:pcl} 
The contracts of seller $\pmv A$ and buyer $\pmv B$ 
from Ex.~\ref{ex:sale:abstract}
can be modelled as follows:
\[
  c_{\pmv A} = {\pmv{A}} \says (({\pmv{B}} \says \atom{pay}) \imp \atom{ship})
  \hspace{50pt}
  c_{\pmv B} = {\pmv{B}} \says \atom{pay}
\]
By the proof system of \pcl\!, we have that: 
\(
    c_{\pmv A} \land c_{\pmv B} \vdash 
    (\pmv{A} \says \atom{ship}) \land 
    (\pmv{B} \says \atom{pay})
\).
\end{example}

\subsection{On contracts-as-processes \emph{vs.} contracts-as-formulae} \label{sec:comparison}

We now compare contracts-as-processes with contracts-as formulae.
We start with an empirical argument, by comparing in 
Table~\ref{tbl:comparison} 
a set of archetypal agreements which use contracts from both paradigms.
Our main technical result is Theorem~\ref{th:pcl-ccs}, where we show that
contracts-as-formulae, expressed in a significant fragment of \pcl\!, 
can be encoded into contracts-as-processes.
Finally, we further discuss the differences between the two contract
models in some specific examples.

\paragraph{Sketching a correspondence.}

We now discuss Table~\ref{tbl:comparison}.
Contracts yielding similar consequences lay on the same row.
Each row tells when {\em interaction} is possible, i.e.~when processes
will eventually reach~$\nil$, and when the formulae entail the observable
$\pmv A \says \atom{a} \,\land\, \pmv B \says \atom{b}$.

In row 1, $\pmv B$ performs $\atom{b}$ unconditionally.
Instead, $\pmv A$ specifies in her contracts a causal dependency
between $\atom{b}$ and $\atom{a}$ --- using a prefix in the world of
processes, or an implication in the world of formulae.  Note that this
exchange offers no protection for $\pmv B$, i.e., $c_{\pmv A}$ could be
replaced with anything else and $\pmv B$ would still be required to
provide $\atom{b}$.

In row 2, $\pmv B$ protects himself by using a causal dependency,
using the dual contract of $c_{\pmv A}$. Now however interaction/entailment
is lost: every principal requires the other one to ``make the first
step'', and circular dependencies forbid any agreement.

In row 3, $\pmv B$ makes the first step by inverting the order of the
causal dependency in $c_{\pmv B}$. The outcome is similar to row 1, except
that now the contract of $\pmv B$ mentions that $\pmv B$ expects
$\atom{a}$ to be performed.  The meaning of the process is roughly
``offer $\atom{b}$ first, then require $\atom{a}$'' which has no
analogous contract-as-formula.

In row 4, $\pmv B$ is offering $\atom{b}$ asynchronously, so removing
the causal dependency. In the world of processes, this is done using
parallel composition instead of a prefix; in the world of (\pcl\!\!)
formulae this is done using $\coimp$ instead of
$\imp$. Interaction/entailment is still possible. 
The contracts $c_{\pmv A}$ and $c_{\pmv B}$ are not symmetric, in which case 
$c_{\pmv A}$ specifies a causal dependency, while $c_{\pmv B}$ does not.

In row 5, causal dependency is removed from $c_{\pmv A}$ as well. 
Both $\pmv A$ and $\pmv B$ are using parallel composition/contractual
implication. This results in symmetric contracts which yield
interaction/entailment.

\begin{table}[t]
\small
\[
\begin{array}{l@{\hspace{10pt}}l@{\hspace{10pt}}l@{\hspace{10pt}}l}
&& \mbox{Contracts-as-processes} & \mbox{Contracts-as-formulae}
\\
\hline
1. &
\begin{array}{l}
c_{\pmv A} \\
c_{\pmv B} \\
{}
\end{array}
&         
\begin{array}{l}
\pmv A \says \atom{b}^-.\atom{a}^0 \\
\pmv B \says \atom{b}^+ \\
\mbox{no protection, interaction}
\end{array}
 &        
\begin{array}{l}
\pmv A \says ({\pmv B} \says \atom{b}) \imp \atom{a} \\
\pmv B \says \atom{b} \\
\mbox{no protection, entailment}
\end{array}
\\ 
\hline
2. &
\begin{array}{l}
c_{\pmv A} \\
c_{\pmv B} \\
{}
\end{array}
&         
\begin{array}{l}
\pmv A \says \atom{b}^-.\atom{a}^+ \\
\pmv B \says \atom{a}^-.\atom{b}^+ \\
\mbox{no interaction}
\end{array}
 &        
\begin{array}{l}
\pmv A \says (\pmv B \says \atom{b}) \imp \atom{a} \\
\pmv B \says (\pmv A \says \atom{a}) \imp \atom{b} \\
\mbox{no entailment}
\end{array}
\\ 
\hline
3. &
\begin{array}{l}
c_{\pmv A} \\
c_{\pmv B} \\
{}
\end{array}
&         
\begin{array}{l}
\pmv A \says \atom{b}^-.\atom{a}^+ \\
\pmv B \says \atom{b}^+.\atom{a}^- \\
\mbox{asymmetric, interaction}
\end{array}
 &        
\begin{array}{l}
\mbox{(no equivalent)}
\end{array}
\\ 
\hline
4. &
\begin{array}{l}
c_{\pmv A} \\
c_{\pmv B} \\
{}
\end{array}
&         
\begin{array}{l}
\pmv A \says \atom{b}^- . \atom{a}^+ \\
\pmv B \says \atom{b}^+ | \atom{a}^- \\
\mbox{asymmetric, interaction}
\end{array}
 &        
\begin{array}{l}
\pmv A \says (\pmv B \says \atom{b}) \imp \atom{a} \\
\pmv B \says (\pmv A \says \atom{a}) \coimp \atom{b} \\
\mbox{asymmetric, entailment}
\end{array}
\\ 
\hline
5. &
\begin{array}{l}
c_{\pmv A} \\
c_{\pmv B} \\
{}
\end{array}
&         
\begin{array}{l}
\pmv A \says \atom{b}^- | \atom{a}^+ \\
\pmv B \says \atom{b}^+ | \atom{a}^- \\
\mbox{symmetric, interaction}
\end{array}
 &        
\begin{array}{l}
\pmv A \says (\pmv B \says \atom{b}) \coimp \atom{a} \\
\pmv B \says (\pmv A \says \atom{a}) \coimp \atom{b} \\
\mbox{symmetric, entailment}
\end{array}
\\ 
\hline
\end{array}
\]
\caption{Contracts-as-processes \emph{vs.}
  contracts-as-formulae} \label{tbl:comparison}
\end{table}

\paragraph{A formal correspondence between contract models.}

We now provide a precise correspondence between two fragments of
contracts-as-formulae and contracts-as-processes, building upon the
intuition underlying the cases shown in Table~\ref{tbl:comparison}.

Concretely, we consider a fragment (\pclminus) of \pcl contracts
comprising atoms, conjunctions, says, and non-nested (contractual)
implications.  Then, we provide an encoding of \pclminus into
contract-as-processes (Def.~\ref{def:pcl-to-ccs}) and formally
relate them.

\begin{definition} \label{def:pclminus}
We define \pclminus as the fragment of \pcl where formulae $c$ have 
the following syntax:
\[
\begin{array}{rll}
c & ::= & \bigwedge_{i \in \mathcal{I}} \; A_i \says \alpha_i \\
\alpha & ::= & \bigwedge_{j \in \mathcal{J}} \; \atom{q}_j 
        \ \ \Big| \ \ 
    ( \bigwedge_{i \in \mathcal{I}} \; B_i \says \atom{p}_i ) 
               \rightarrow \bigwedge_{j \in \mathcal{J}} \atom{q}_j
        \ \ \Big| \ \ 
   ( \bigwedge_{i \in \mathcal{I}} \; B_i \says \atom{p}_i ) 
               \coimp \bigwedge_{j \in \mathcal{J}} \atom{q}_j
\end{array}
\]
\end{definition}

\begin{definition} \label{def:pcl-to-ccs}
For all formulae $c$ of \pclminus, we define 
the contract-as-process $[c]$ as follows,
where, for all atoms $\atom{q}$, 
$\mathit{OUT}(\atom{q})$ is a recursive process defined by the equation
$\mathit{OUT}(\atom{q}) = \tau^0.\nil + \atom{q}^+ . \mathit{OUT}(\atom{q})$.
Also, we assume an injective function $\__{/\_}$ that maps 
all atoms $\atom{q}$ and all principals $A$ into
the atom $\atom{q}_{/A}$.
\[
\begin{array}{lll}
[ \bigwedge_{i \in \mathcal{I}} \; A_i \says \alpha_i ] & = & 
   ||_{i \in \mathcal{I}} \; A_i \says [ {\alpha}_i ]_{A_i} \\ 
{}[ \bigwedge_{j \in \mathcal{J}} \; \atom{q}_j ]_{A} & = & 
       ||_{j \in \mathcal{J}} \; \mathit{OUT}(\atom{q}_{j/A}) \\ 
{}[ ( \bigwedge_{i\in\setenum{1,\ldots,n}} \; B_i \says \atom{p}_i ) 
               \rightarrow \bigwedge_{j \in \mathcal{J}} \atom{q}_j ]_{A}
    & = & 
    \atom{p}_{1/{B_1}}^- .\, \cdots .\, \atom{p}_{n/B_n}^- .\,
    \Big( ||_{j \in \mathcal{J}} \; \mathit{OUT}(\atom{q}_{j/A}) \Big) \\ 
{}[ ( \bigwedge_{i\in\setenum{1,\ldots,n}} \; B_i \says \atom{p}_i ) 
               \coimp \bigwedge_{j \in \mathcal{J}} \atom{q}_j ]_{A} & = &
    \Big( ||_{j \in \mathcal{J}} \; \mathit{OUT}(\atom{q}_{j/A}) \Big) 
    \ \Big| \ 
    \atom{p}_{n/B_1}^- .\, \cdots .\, \atom{p}_{n/B_n}^- .\, \nil
\end{array}
\]
\end{definition}

In Def.~\ref{def:duties} below we extract from a \pclminus contract
$c$ the associated {\it latent actions}, i.e.~the set of all
atoms occurring in $c$ paired with the participant who may have to
abide by.

\begin{definition} \label{def:duties}
The function $\lambda$ from \pclminus formulae to
sets of actions is defined as follows:
\begin{align*}
  & \textstyle \lambda( \bigwedge_{i \in \mathcal{I}} \; A_i \says {\alpha}_i ) 
  = 
  \bigcup_{i \in \mathcal{I}} \lambda_{A_i}( {\alpha}_i ) 
  \\
  & \textstyle \lambda_{A}( \bigwedge_{j \in \mathcal{J}} \; \atom{q}_j ) 
  =
  \setcomp{ A \says \atom{q}_j }{ j \in \mathcal{J} }
  \\
  & \textstyle \lambda_{A}( ( \bigwedge_{i \in \mathcal{I}} \; B_i \says \atom{p}_i ) 
               \imp \bigwedge_{j \in \mathcal{J}} \atom{q}_j ) 
  = 
  \setcomp{A \says \atom{q}_j}{j \in \mathcal{J}}
  \cup
  \setcomp{B_i \says \atom{p}_i}{i \in \mathcal{I}} \\
  & \textstyle \lambda_{A}( ( \bigwedge_{i \in \mathcal{I}} \; B_i \says \atom{p}_i ) 
               \coimp \bigwedge_{j \in \mathcal{J}} \atom{q}_j ) 
  = 
  \setcomp{A \says \atom{q}_j}{j \in \mathcal{J}}
  \cup
  \setcomp{B_i \says \atom{p}_i}{i \in \mathcal{I}}
\end{align*}
\end{definition}

The following result establishes a correspondence between our contract
models. A \pclminus contract $c$ requires to perform all the latent
actions if\cem{ and only if}{f} its encoding $[c]$ can be fulfilled.

\begin{theorem}\label{th:pcl-ccs}
For all \pclminus formulae $c$ involving principals $\{A_i\}_{i \in I}$,
the following are equivalent:
\begin{itemize}
\item 
(contracts-as-formulae)
\(
  \forall c',\mu_1,\ldots,\mu_n . \ 
  \big(
  ( c \xrightarrow{\mu_1}\cdots \xrightarrow{\mu_n} c' \;\land\;
  \forall i \in I . \ \honest{c'}{A_i} )
  \implies 
  \lambda(c) \subseteq \setenum{\mu_1,\ldots,\mu_n}
  \big)
\)

\item 
(contracts-as-processes)
\(
  \exists c'. \ 
  (
  [ c ] \xrightarrow{}^* c'
  \;\land\;
  \forall i \in I. \
  \honest{c'}{A_i}
  )
\)
\end{itemize}
\end{theorem}

The above statement can in fact be reduced to the result below by
considering the definition of $\honest{}{}$ in both contract models.

\begin{theorem} 
For all \pclminus formulae $c$, $c \vdash \lambda(c)$ if and only if
$[ c ] \xrightarrow{}^* \nil$.
\end{theorem}

The ``if'' direction mainly follows from the fact that, unless a
$\tau^0$ is fired, the residuals of $[c]$ are of the form $[c']$ where
$c'$ is a formula equivalent to $c$.
The ``only if'' direction is proved by considering the order of
applications of modus ponens in a proof of $c \vdash \lambda(c)$ and
consequently firing the inputs corresponding to premises of
implications. This makes their consequences/outputs
available. The case of contractual implications is simpler because
outputs are immediately available, hence their generated inputs
can be fired last.


\section{A basic calculus for contract-oriented computing}\label{sec:co2}

We now introduce the syntax and semantics of \coco\!.  
It generalises the contract calculus in~\cite{BZ10lics} by making
it independent of the actual contract language.
In fact, \coco assumes the abstract model of contracts
introduced in \S~\ref{sec:contract-model}, which can
be instantiated to a variety of actual contract languages.
While taking inspiration from Concurrent Constraint Programming,
\coco makes use of more concrete
communication primitives which do not assume a global constraint
store, so reducing the gap towards a possible distributed
implementation.
The main differences between \coco and CCP are:
($i$) in \coco constraints are multisets of contracts,
($ii$) in \coco there is no global store of constraints: all the prefixes act on sessions,
($iii$) the prefix $\ask{}{}\!$ of \coco also instantiates variables to names, and
($iv$) the prefixes $\fact{}{}\!$ (which makes a multiset of constraints evolve) 
and $\fuse{}{}\!$ (which establishes a new session) have no counterpart in CCP.

\subsection{Syntax}

First, let us define the syntax of \coco\!.

\begin{definition} \label{def:calculus:syntax}
The abstract syntax of \coco is given by the following productions:
\[\begin{array}{rcccccccccccc}
    \text{Systems}   \quad S  & ::= & 
    \emptysys &\sep& \sys {\pmv A} P &\sep& \sys s C &\sep& S \mid S &\sep& (u)S
    \\[.3pc] 
    \text{Processes} \quad P  & ::=  & \freeze u c
    &\sep&    \textstyle \sum_{i} \pi_i.P_i
    &\sep&    P \mid P
    &\sep&    (u)P 
    &\sep&    X(\vec u)
    \\[.3pc] 
    \text{Prefixes}  \quad \pi
    & ::= & \tau
    &\sep&    \fact u {\atom{a}}
    &\sep&    \tell u {\freeze v c}
    &\sep&    \ask {u,\vec{v}} \phi
    &\sep&    \fuse u \phi 
\end{array}\]
We stipulate that the following conditions always hold:
in a system $||_{i \in \mathcal{I}} \sys{n_i}{P_i}$, $n_i
\neq n_j$ for each $i \neq j \in I$;
in a process $(u)P$ and a system $(u)S$, $u \not\in \pnames$.
\end{definition}

We distinguish between \emph{processes} and \emph{systems}.
Systems $S$ consist of a set of \emph{agents} $\pmv A[P]$
and of \emph{sessions} $s[C]$, composed in parallel.
Processes $P$ comprise latent contracts, 
guarded summation, parallel composition, scope delimitation,
and process identifiers.
A \emph{latent contract} $\freeze x c$ represents a contract $c$ which 
has not been stipulated yet; upon stipulation, the variable $x$ 
will be bound to a fresh session name.
We allow finite prefix-guarded sums of processes; as usual,
we write $\pi_1.P_1 + \pi_2.P_2$ for $\sum_{i=1,2}\pi_i.P_i$.
Processes can be composed in parallel, and can be put under the scope of
binders $(u)\_$.
We use process identifiers $X$ to express recursive processes, and we
assume that each identifier has a corresponding defining equation
$X(u_1,\ldots,u_j) \mmdef P$ such that $\fv{P} \subseteq
\{u_1,\ldots,u_j\} \subseteq \vars$ and each occurrence of 
process identifiers $Y$ in $P$
is prefix-guarded.
Note that principal names cannot be bound, but can be communicated if
permitted by the contract language.
\cem{}{The empty system and the empty sum are both denoted by $\emptysys$.
As usual, we may omit trailing occurrences of $\nil$ in processes.
Free/bound occurrences of variables/names are defined as expected.
Variables and session names can be bound;
in $(u)S$ all free occurrences of $u$ in $S$ are bound;
$S$ is \emph{closed} when it has no free variables.  } 
Prefixes include the silent prefix $\tau$, 
action execution $\fact{u}{\atom a}$, 
contract advertisement $\tell{u}{\freeze v c}$, 
contract query $\ask{u,\vec{v}}{\phi}$, 
and contract stipulation $\fuse{u}{\phi}$. 
The index $u$ indicates the target agent/session
where the prefix will be fired\cem{.}{{} or the session where
stipulations are kept, in the case of $\fuse{}{}$.}
We shall refer to the set of latent contracts within an agent ${\pmv
  A}[P]$ as its {\em environment}; similarly, in a session $s[C]$ we
shall refer to $C$ as the environment of $s$.
Note that the environment of agents can only contain latent contracts,
advertised through the primitive $\tell{}{}\!$, while sessions
can only contain contracts, obtained either upon reaching an
agreement with $\fuse{}{}\!$, or possibly upon principals performing
some $\fact{}{}$ prefixes.

\newcommand{\redar}{\ \to\ }

\subsection{Semantics}
  
\begin{table}[t]
  \hrulefill
  \small
\begin{center}
commutative monoidal laws for $\mid$ on processes and systems
\end{center}
\vspace{-20pt}
\begin{align*}
  \sys u {(v)P} & \equiv \sys{(v)u} P 
  \quad \text{if}\ u \neq v
  \hspace{0pt}
  & Z \mid (u)Z' & \equiv (u)(Z \mid Z') 
  \quad \text{if}\ u \not\in \fv Z \cup \fn Z
  \\[8pt]
  (u)(v)Z & \equiv (v)(u)Z
  \hspace{0pt}
  & (u)Z & \equiv Z \quad \text{if}\ u \not\in \fv Z \cup \fn Z
\end{align*}
\vspace{-10pt}
\[
  \freeze n c \equiv \nil
  \hspace{25pt}
  \tell u {\freeze n c}.P \equiv \nil 
  \hspace{25pt}
  \ask {u,\vec{v}} \phi. P \equiv \nil  
  \quad \text{if}\ \vec{v} \cap \names \neq \emptyset
  \hspace{25pt} 
  \fuse n \phi. P \equiv \nil
\]
  \hrulefill
  \vspace{-0pt}
\caption{Structural equivalence ($Z,Z'$ range over systems or
  processes)} \label{tbl:equiv}
\end{table}

The semantics of \coco is formalised by a reduction relation on
systems which relies on the structural congruence laws in
Table~\ref{tbl:equiv}.
Only the last row in Table~\ref{tbl:equiv} contains non-standard laws:
they allow for collecting garbage terms which may possibly arise
after variable substitutions.

\begin{definition}\label{def:calculus:semantics}
  The binary relation $\redar$ on closed systems is the smallest
  relation closed under structural congruence and under the rules in
  Table~\ref{tbl:semantics}, where the relation $\agreement K \phi x
  \sigma$ \cem{}{in \rname{Fuse}} is introduced in
  Def.~\ref{def:agreement}, and $\thaw K$ is obtained by removing all
  the $\freeze \_ {}$ from $K$, i.e.\ if 
  $K = ||_{i \in I} \freeze{x_i}{c_i}$, then 
  $\thaw K = ||_{i \in I} \, c_i$.  
  Also, we identify the parallel composition of contracts $C =
  c_1 \mid \ldots \mid c_j$ with the multiset
  $\setenum{c_1,\ldots,c_j}$ (similarly for latent contracts).
\end{definition}

\begin{table}
\hrulefill
  \[
  \begin{array}{cr}
    {\sys {\pmv A} {\tau . P + P' \mid Q}
      \xrightarrow{}
      \sys {\pmv A} {P \mid Q}
    } & \rname{Tau}
    \\[20pt]
    {\sys {\pmv A} {\tell {\pmv A} {\freeze x c} . P + P' \mid Q}
      \xrightarrow{}
      {\sys {\pmv A} {\freeze x {\pmv A \says c} \mid P \mid Q}}
    } & \rname{Tell$_1$}
    \\[20pt]
    {\sys {\pmv A} {\tell {\pmv B} {\freeze x c} . P + P' \mid Q} \ \mid\ \sys {\pmv B} R
      \xrightarrow{}
      \sys {\pmv A} {P \mid Q} \ \mid\ \sys {\pmv B} {R \mid \freeze x {\pmv A \says c}}
    } & \rname{Tell$_2$}
    \\[20pt]
    \redrule
    {C \xrightarrow{\tuple{\pmv A_1 \says \atom{a}_1, \ldots, \pmv A_j \says \atom{a}_j}} C'
      \qquad j \geq 1
    }
    {\sys s {C} \ \mid \ \ \dmid_{1\leq i\leq j}
      \sys {\pmv A_i} {\fact s {\atom{a}_i} . P_i + P_i' \mid Q_i}
      \xrightarrow{} 
      \sys s {C'} \ \mid\ \ \dmid_{1\leq i\leq j}
      \sys {\pmv A_i} {P_i \mid Q_i}
    } & \rname{Do}
    \\[30pt]
    \redrule
    {\dom \sigma = \vec{u} \subseteq \vars \qquad C\sigma \vdash \phi\sigma}
    {(\vec u)(\sys {\pmv A} {\ask {s,\vec{u}} \phi . P + P'  \mid Q} \mid \sys s C \mid S)
      \xrightarrow{}
      \sys {\pmv A} {P \mid Q}\sigma \ \mid\ \sys s C\sigma \mid S\sigma
    } & \rname{Ask}
    \\[30pt]
    \redrule
    {\agreement K \phi x \sigma
      \qquad \vec u = \dom \sigma \subseteq \vars
      \qquad s = \sigma(x) \;\; \text{fresh}
    }
    {(\vec u)(\sys {\pmv A} {\fuse x \phi . P + P'  \mid K \mid Q} \mid S)
      \xrightarrow{}
      (s)(\sys {\pmv A} {P \mid Q}\sigma \ \mid\ \sys s {\thaw K}\sigma \mid S\sigma)
    } & \rname{Fuse}
    \\[30pt]
    \redrule
    {X(\vec u) \mmdef P \qquad P\subs{\vec v}{\vec u} \xrightarrow{} P'}
    {X(\vec v) \xrightarrow{} P'}
    & \rname{Def} \\[30pt]
    \redrule
    {S \xrightarrow{} S'}
    {S \mid S'' \xrightarrow{} S' \mid S''}
    & \rname{Par} \\[30pt]
    \redrule
    {S \xrightarrow{} S'}
    {(u)S \xrightarrow{} (u)S'}
    & \rname{Del}
  \end{array}
  \]
\hrulefill
\caption{Reduction semantics of \coco} \label{tbl:semantics}
\end{table}

Axiom \rname{Tau} and rules \rname{Par}, \rname{Del}, and \rname{Def}
are standard.
Axioms \rname{Tell$_1$} and \rname{Tell$_2$} state that a principal
\pmv A can advertise a latent contract $\freeze x c$ either in her own
environment, or in a remote one.
In \rname{Do}, the principals $\pmv A_i$ are participating in 
the contracts $C$ stipulated on session $s$. 
Basically,  the system evolves when all the involved principals 
are ready to fire the required action.
Rule \rname{Ask} allows $\pmv A$ to check if an observable $\phi$ is
entailed by the contracts $C$ in session $s$; notice that the
entailment is subject to an instantiation $\sigma$ of the variables
mentioned in the ask prefix.
Rule \rname{Fuse} establishes a multi-party session $s$ among all the parties
that reach an agreement on the latent contracts $K$.
Roughly, the {\em agreement} relation $\agreement K \phi x \sigma$ 
(Def.~\ref{def:agreement}) 
holds when, upon some substitution
$\sigma$, the latent contracts $K$ entail $\phi$, and the fresh
session name $s$ substitutes for the variable $x$.

The simplest typical usage of these primitives is as follows. First, a
group of principals exchanges latent contracts using $\tell{}{}$,
hence sharing their intentions. Then, one of them opens a new session
using the $\fuse{}{}$ primitive. Once this happens, each involved
principal $\pmv A$ can inspect the session using $\ask{}{}$, hence
discovering her actual duties within that session: in general, these
depend not only on the contract of $\pmv A$, but also on those of the
other principals (see e.g.~Ex.\ref{ex:compliance}). Finally, primitive
$\fact{}{}$ is used to actually perform the duties.

\begin{example} \label{ex:sale:pcl-co2} The sale scenario 
between seller $\pmv A$ and buyer $\pmv B$ from Ex.~\ref{ex:sale:pcl}
can be formalized as follows.
\[
\begin{array}{lcl}
  S \;\; &=& \;\;
  \sys {\pmv{A}}
  {
    (x,b) \;
    \tell {\pmv{A}} {\freeze x {((b \says \atom{pay}) \imp \atom{ship})}} . \;
    \fuse x {(\pmv{A} \says \atom{ship})} . \;
    \fact x {\atom{ship}} 
  }
\\
  & \mid & \;\;
  \sys {\pmv{B}}
  { 
    (y) \;
    \tell {\pmv{A}} {\freeze y {\atom{pay}}} . \;
    \ask y {(\pmv{B} \says \atom{pay})} . \;
    \fact y {\atom{pay}} 
  }
\end{array}
\]
The buyer tells $\pmv{A}$ a contract which binds \pmv B to pay,
then waits until discovering that he actually has to pay,
and eventually does that.
A session between the buyer and the seller is created and proceeds
smoothly as expected:
\[
\begin{array}{rcl}
  S
  & \longrightarrow &
  (x,b,y) \;
  \big(
  \sys {\pmv{A}}
  {
    {\freeze x {\pmv A \says ((b \says \atom{pay}) \imp \atom{ship})}} \; | \;
    \fuse x {(\pmv{A} \says \atom{ship})} . \;
    \fact x {\atom{ship}} 
  }
\\
  &&
  \hspace{36pt} | \; \sys {\pmv{B}}
  { 
    \tell {\pmv{A}} {\freeze y {\atom{pay}}} . \;
    \ask y {(\pmv{B} \says \atom{pay})} . \;
    \fact y {\atom{pay}} 
  }
  \; \big)
\\[7pt]
  & \longrightarrow &
  (x,b,y) \;
  \big(
  \sys {\pmv{A}}
  {
    {\freeze y {\pmv B \says \atom{pay}}} \; | \;
    {\freeze x {\pmv A \says ((b \says \atom{pay}) \imp \atom{ship})}} \; | \;
    \fuse x {(\pmv{A} \says \atom{ship})} . \;
    \fact x {\atom{ship}} 
  }
\\
  &&
  \hspace{36pt} | \; \sys {\pmv{B}}
  { 
    \ask y {(\pmv{B} \says \atom{pay})} . \;
    \fact y {\atom{pay}} 
  }
  \; \big)
\\[7pt]
  & \longrightarrow &
  (s) \;
  \big(
  \sys {\pmv{A}}
  {
    \fact s {\atom{ship}} 
  }
  \; | \; \sys {\pmv{B}}
  { 
    \ask s {(\pmv{B} \says \atom{pay})} . \;
    \fact s {\atom{pay}} 
  }
  \; | \;
  \sys s { \pmv A \says ((\pmv B \says \atom{pay}) \imp \atom{ship})
         \ ,\  \pmv B \says {\atom{pay}}
         }
  \big)
\\[7pt]
  & \longrightarrow &
  (s) \;
  \big(
  \sys {\pmv{A}}
  {
    \fact s {\atom{ship}} 
  }
  \; | \; \sys {\pmv{B}}
  { 
    \fact s {\atom{pay}} 
  }
  \; | \;
  \sys s { \pmv A \says ((\pmv B \says \atom{pay}) \imp \atom{ship})
         \ ,\  \pmv B \says {\atom{pay}}
         }
  \; \big)
\\[7pt]
  & \longrightarrow &
  (s) \;
  \big(
  \sys {\pmv{A}}
  {
  \nil
  }
  \; | \; \sys {\pmv{B}}
  { 
    \fact s {\atom{pay}} 
  }
  \; | \;
  \sys s { \pmv A \says ((\pmv B \says \atom{pay}) \imp \atom{ship})
         \ ,\  \pmv B \says {\atom{pay}}
         \ ,\  \pmv A \says {!\atom{ship}} 
         \ ,\  \ldots
         }
  \; \big)
\\[7pt]
  & \longrightarrow &
  (s) \;
  \big(
  \sys {\pmv{A}}
  {
  \nil
  }
  \; | \; \sys {\pmv{B}}
  { 
  \nil
  }
  \; | \;
  \sys s { \pmv A \says ((\pmv B \says \atom{pay}) \imp \atom{ship})
         \ ,\  \pmv B \says {\atom{pay}}
         \ ,\  \pmv A \says {!\atom{ship}} 
         \ ,\  \pmv B \says {!\atom{pay}} 
         \ ,\  \ldots
         }
  \; \big)
\end{array}
\]
\end{example}

In the previous example, we have modelled the system outlined 
in Ex.~\ref{ex:sale:abstract} using a contracts-as-formulae approach.
In the following example, we adopt instead the contracts-as-processes
paradigms.
In the meanwhile, we introduce a further participant to our system:
a broker \pmv C which collects the contracts from \pmv A and \pmv B,
and then uses a $\fuse{}{}$ to find when an agreement is possible.

\begin{example} \label{ex:sale:ccs-co2}
Recall the contract of Ex.~\ref{ex:sale:ccs}.
We specify the behaviour of the system as:
\[\begin{array}{rcl}
    S & \mmdef & \pmv A[(x) (\tell {\pmv C} {(\freeze {x} {\atom{pay}^-.\atom{ship}^0})}.\ \ask{x}{\Diamond{\atom{pay}^+}}.\ \fact{x}{\atom{pay}^-}.\ \fact{x}{\atom{ship}^0}]
    \\
    & \mid &     \pmv B[(y) (\tell {\pmv C} {\freeze {y} {\atom{pay}^+}}.\ \fact{y}{\atom{pay}^+} ]
    \\
    & \mid &     \pmv C[(z)(\fuse z \;\Diamond(\atom{pay}^+ \land \Diamond\atom{ship}^0))]
  \end{array}
\]
The principals $\pmv A$ and $\pmv B$ advertise their contracts to
the broker $\pmv C$, which opens a session for their
interaction.
As expected, as soon as the payment is received, the goods are
shipped.
After advertising such contract in her
environment, the seller waits until finding that she has promised to
ship; after that, she actually ships the item.  
\[
  S
  \;\; \longrightarrow^* \;\;
  (s) \;
  \big(
  \sys{\pmv{A}} {\zero} \;\mid\;
  \sys{\pmv{B}} {\zero} \;\mid\;
  \sys{s} { \nil }
  \big)
\]
\end{example}

\subsection{On agreements}

To find agreements ($\agreement K \phi x \sigma$), 
we use the relation $\vdash$ of the contract model.

\vbox{
\begin{definition}\label{def:agreement}
  For all multisets $K$ of latent contracts, for all substitutions $\sigma: \vars \to
  \names$, for all $x \in \vars$, for all observables $\phi$, 
  we write $\agreement K \phi x \sigma$ iff the following conditions hold:
  \begin{itemize}
  \item $x \in \dom \sigma$
  \item $\fv{K\sigma} = \fv{\phi\sigma} = \emptyset$
  \item $\exists s \in \snames: \forall y \in \dom \sigma \cap \svars: \sigma(y) = s$
  \item $(\thaw K)\sigma \entails \phi\sigma$
  \item no $\sigma' \subset \sigma$ satisfies the conditions above,
    i.e.\ $\sigma$ is \emph{minimal}.
  \end{itemize}
\end{definition}}

Basically, Def.~\ref{def:agreement} states that an agreement is
reached when the latent contracts in $K$ entail, under a suitable
substitution, an observable $\phi$.
Recall that $\phi$ is the condition used by the principal acting as
broker when searching for an agreement (cf. rule \rname{Fuse} in
Table~\ref{tbl:semantics}).
A valid agreement has to instantiate with a minimal substitution
$\sigma$ all the variables appearing in the latent contracts in $K$ as
well as the session variable $x$; the latter, together with any other
session variables in the domain of the substitution $\sigma$, is
mapped to a (fresh) session name $s$.

The minimality condition on $\sigma$ forces brokers to include in
agreements only ``relevant'' participants.
For instance, let $\freeze{x_1}{c_1}$, $\freeze{x_2}{c_2}$, and
$\freeze{x_3}{c_3}$ be the latent contracts advertised to a broker; if
there is an agreement on $\freeze{x_1}{c_1}$ and $\freeze{x_2}{c_2}$,
the latent contract $\freeze{x_3}{c_3}$ would not be included.
Basically, the minimality condition allows brokers to tell apart
unrelated contracts.
This is illustrated by the following informal example.

\begin{example}
Let $\phi$ be the observable ``$\pmv A$ shall ship the goods'',
and consider the following latent contracts,
advertised by $\pmv A$, $\pmv B$, and $\pmv C$, respectively:
\begin{itemize}
\item $\freeze{x_1}{c_1} =$ ``in session $x_1$, if some principal $b$ pays, 
        then I shall ship the goods''
\item $\freeze{x_2}{c_2} =$ ``in session $x_2$, I shall pay''
\item $\freeze{x_3}{c_3} =$ ``in session $x_3$, I shall kiss a frog''.
\end{itemize}
Note that the first two latent contracts do entail $\phi$ when
  $\sigma(b) = \pmv B$, and $\sigma(x_1) = \sigma(x_2) = s$.
  Without the minimality condition, $\sigma(x_3) = s$ would
  have been included in the agreement of $\pmv A$ and $\pmv B$, despite
  the offer of $\pmv{C}$ being somewhat immaterial.
\end{example}

\begin{example}\label{ex:compliance}
Our approach allows for contract models with multiple levels of
compliance. For instance, let $c_{\pmv{A}}$, $c_{\pmv{B}}$, and
$c_{\pmv{B'}}$ be the following PCL contracts:
\[
  c_{\pmv{A}} = \freeze {x_1} {\pmv A} \says 
  \big( (b \says \atom{b}) \imp \atom{a} \;\land\; 
    (b \says \atom{b'}) \imp \atom{a'} \big)
  \qquad
  c_{\pmv{B}} = \freeze {x_2} {\pmv B} \says \atom{b}
  \qquad
  c_{\pmv{B'}} = \freeze {x_3} {\pmv B'} \says \atom{b} \land \atom{b'}
\]
The contract $c_\pmv{A}$ is intuitively compliant with both
$c_{\pmv{B}}$ and $c_{\pmv{B'}}$. When coupled with $c_{\pmv{B'}}$,
the contract $c_{\pmv{A}}$ entails both the obligations $\atom{a}$
and $\atom{a'}$ for $\pmv A$. 
Conversely, when coupled with $c_{\pmv{B}}$, we
obtain a weaker agreement, since only the obligation $\atom a$ is
entailed. 
Although both levels of agreement are possible,
in some sense the contract $c_{\pmv{B'}}$ provides
$c_{\pmv{A}}$ with a better Service-Level Agreement than $c_{\pmv{B}}$. 
Through the primitive $\ask{}{}$, principal $\pmv A$ can detect the
actual service level she has to provide.
\end{example}

\subsection{On violations} \label{sec:violations}

In Def.~\ref{def:culpable} below we set out when a principal $\pmv A$ 
is honest in a given system $S$.
Intuitively, we consider all the possible runs of $S$, and require
that in every session the principal $\pmv A$ eventually fulfils all
her duties.
To this aim, we shall exploit the fulfilment relation $\honest{C}{\pmv
  A}$ from the contract model.

To do that, we need to cope with a few technical issues. First, the
$\alpha$-conversion of session names makes it hard to track the {\em
same} session at different points in a trace. So, we consider
traces ``without $\alpha$-conversion of names''.
Technically, let
\[
\mathit{freezeNames}(S) = \setcomp{S'}{S\equiv (s_1, \ldots, s_j) S' 
  \text{ and }
  S' \mbox{ free from $(s)$ delimitations}}
\]
and define $S \rightsquigarrow S''$ whenever $S \xrightarrow{} S'$ and $
S'' \in \mathit{freezeNames}(S')$.
A $\rightsquigarrow$-trace is a trace w.r.t.~$\rightsquigarrow$. Such a 
$\rightsquigarrow$-trace is
\emph{maximal} iff it is either infinite or ending with $S_j
\not\rightsquigarrow$.

Another technical issue is that a principal could not get a chance to
act in all the traces.
For instance, consider the system $S = \sys {\pmv A}{\fact s
  {\atom{pay}}} \mid \sys {\pmv B}{X} \mid S'$ where $S'$ enables \pmv
A's action and $X \defeq \tau.X$; note that $S$ generates the infinite
trace $S \rightsquigarrow S \rightsquigarrow S \rightsquigarrow \cdots$ in
which \pmv A never pays, despite her honest intention.
To account for this fact, we will check the honesty of a principal in
{\em fair} traces, only, i.e.~those obtained by running $S$ according
to a fair scheduling algorithm.
More precisely, we say that $\rightsquigarrow$-trace
$(S_i)_i$ is {\em fair} if no
single prefix can be fired in an infinite number of $S_i$.  

\begin{definition} \label{def:culpable}
  A principal $\pmv A$ is \emph{honest} in $S$ iff for all maximal
  fair $\rightsquigarrow$-traces $(S_i)_{i\in \mathcal{I}}$,
  \[
  \forall s.\;\;
  \exists j\in \mathcal{I}.\;\;
  \forall i \geq j,\; \vec{n},\; C,\; S'.\;\; 
  \Big(
  S_i \equiv (\vec{n}) ( s[C] \ \mid\ S' )
  \implies
  \honest{C}{\pmv A} 
  \Big)
  \]
\end{definition}
In other words, a principal \pmv A misbehaves
if involved in a session $s$ such that the contracts $C$ of $s$ 
do not settle the obligations of \pmv A.

\begin{example} \label{ex:sale:violation} 
Consider the variation of Ex.~\ref{ex:sale:pcl-co2},
where the seller is modified as follows:
\[
\begin{array}{lcl}
  S \;\; &=& \;\;
  \sys {\pmv{A}}
  {
    (x,b) \;
    \tell {\pmv{A}} {\freeze x {((b \says \atom{pay}) \imp \atom{ship})}} . \;
    \fuse x {(\pmv{A} \says \atom{ship})} . \;
    \fact x {\atom{snakeOil}} 
  } 
\\[5pt]
  & \mid & \; 
  \sys {\pmv{B}}
  { 
    (y) \;
    \tell {\pmv{A}} {\freeze y {\atom{pay}}} . \;
    \ask y {(\pmv{B} \says \atom{pay})} . \;
    \fact y {\atom{pay}} 
  }
\end{array}
\]
The fraudulent seller $\pmv{A}$ promises to ship ---
but eventually only provides the buyer with some snake oil.
The interaction between $\pmv{A}$ and $\pmv{B}$ leads to a violation:
\begin{align*}
  S
  \;\; \rightarrow^* \;\;
  (s) \;
  \big(
  \sys{\pmv{A}} {\zero} \;\mid\;
  \sys{\pmv{B}} {\zero} \;\mid\;
  s[
    & \pmv{A} \says ((\pmv{B} \says \atom{pay}) \imp \atom{ship}),\;
    \pmv{A} \says !\atom{snakeOil},\; \\
    & \pmv{B} \says \atom{pay},\;
    \pmv{B} \says !\atom{pay} ,\; \ldots
  ]
  \big)
\end{align*}
The buyer has not obtained what he has paid for. Indeed, the seller is
dishonest according to Def.~\ref{def:culpable}, because the contracts
in $s$ entail the promise $\pmv{A} \says \atom{ship}$, which is not
fulfilled by any fact.  A judge may thus eventually punish $\pmv{A}$
for her misconduct.
\end{example}

\subsection{On protection}

We now illustrate some examples where one of the parties is fraudulent.

\begin{example} \label{ex:sale:2} 
  Recall Ex.~\ref{ex:sale:violation} and consider a fraudulent seller
  $\pmv{A}$, which promises in her contract some snake oil.
  The buyer $\pmv B$ is unchanged from that example.
\[
  S \;\; = \;\;
  \sys {\pmv{A}}
  {
    (x,b) \;
    \tell {\pmv{A}} {\freeze x {((b \says \atom{pay}) \imp \atom{snakeOil})}} . \;
    \fuse x {(\pmv{A} \says \atom{snakeOil})} . \;
    \fact x {\atom{snakeOil}} 
  }
  \mid \sys{\pmv B} {\ldots}
\]
The interaction between $\pmv A$ and $\pmv B$ now goes unhappily for
$\pmv B$: he will pay for some snake oil, and $\pmv A$ is not even
classified as dishonest according to Def.~\ref{def:culpable}: 
indeed, $\pmv{A}$ has eventually fulfilled all her promises.
\[
  S
  \;\; \longrightarrow^* \;\;
  (s) \;
  \big(
  \sys{\pmv{A}} {\zero} \;\mid\;
  \sys{\pmv{B}} {\zero} \;\mid\;
  \sys{s} {
    \pmv{A} \says !\atom{snakeOil},\; 
    \pmv{B} \says !\atom{pay} ,\; \ldots }
  \big)
\]
\end{example}

\begin{example} \label{ex:sale:3}
To protect the buyer from the fraud outlined in Ex.~\ref{ex:sale:2}, 
we change the contract of the buyer $\pmv{B}$ as follows:
\[
\begin{array}{lcl}
  S \;\; &=& \;\;
  \sys {\pmv{A}}
  {
    (x,b) \;
    \tell {\pmv{A}} {\freeze x {((b \says \atom{pay}) \imp \atom{snakeOil})}} . \;
    \fuse x {(\pmv{A} \says \atom{snakeOil})} . \;
    \fact x {\atom{snakeOil}} 
  } 
\\[5pt]
  & \mid & \; 
  \sys {\pmv{B}}
  { 
    (a,y) \;
    \tell {\pmv{A}} {\freeze y {(a \says \atom{ship}) \coimp \atom{pay}}} . \;
    \ask y {(\pmv{B} \says \atom{pay})} . \;
    \fact y {\atom{pay}} 
  }
\end{array}
\]
Note that we have used contractual implication $\coimp$,
rather than standard implication $\imp$,
which allows $\pmv{B}$ to reach an agreement also
with the {\em honest} seller contract $(b \says \atom{pay}) \imp \atom{ship}$.
Instead, the interaction between the above {\em fraudulent} seller
$\pmv{A}$ and the buyer $\pmv{B}$ will now get stuck on the
$\fuse{}{}$ in $\pmv{A}$, because the available latent contracts do
not entail $\pmv{A} \says \atom{snakeOil}$.
\end{example}

When using contracts-as-processes, a broker can participate in
deceiving a principal.

\begin{example}\label{ex:ecomm0}
  Consider a simple e-commerce scenario:
  \[\begin{array}{rcl}
    S & \mmdef & \pmv A_1[(x) \tell {\pmv B} {\freeze {x} {(\atom{pay}^+.\atom{ship}^-)}}.\ \fact{x}{\atom{pay}^+}.\ \fact{x}{\atom{ship}^-}]
    \\
    & \mid &     \pmv A_2[(y) \tell {\pmv B} {\freeze {y} {(\atom{pay}^-. (\atom{ship}^+ + \atom{fraud}^0))}}.\ \fact{y}{\atom{pay}^-}.\ \fact{y}{\atom{fraud}^0} ]
    \\
    & \mid &     \pmv B[(z) \fuse z \;\Diamond(\atom{ship}^+ \lor \atom{fraud}^0)]
  \end{array}
  \]
\end{example}
Above, the broker $\pmv B$ and $\pmv A_2$ dishonestly cooperate and
open a session to swindle $\pmv A_1$.
The principal~$\pmv A_2$ will be able to
fulfil her contract ($\honest{C}{\pmv A_2}$), while $\pmv A_1$ will
never receive her goods.
Nevertheless, $\pmv A_1$ is considered culpable, because he cannot perform
the promised action $\atom{ship}^-$.
In Sect.~\ref{sec:variants} we propose a variant of the
contracts-as-processes model allowing principals to
protect from such kind of misbehavior.

\begin{example}\label{ex:ecomm1}
  Consider the following formalization of the e-commerce scenario:
  \[\begin{array}{rcl}
    S & \mmdef & \pmv A_1[(x,a_2) \tell {\pmv B} {\freeze {x} {(a_2 \says \atom{ship} \coimp \atom{pay})}}.\ \ask{x}{\pmv A_1 \says \atom{pay}}.\ \fact{x}{\atom{pay}}]
    \\
    & \mid &     \pmv A_2[(y,a_1) \tell {\pmv B} {\freeze {y} {(a_1 \says \atom{pay} \coimp (\atom{ship} \lor \atom{fraud}))}}.\ \fact{y}{\atom{fraud}} ]
    \\
    & \mid &     \pmv B[(z) \fuse z \phi]
  \end{array}
  \]
  Here, choose $\phi$ so to cause $\pmv A_1$ and $\pmv A_2$ to initiate a session
  (the actual formula is immaterial).
\end{example}

In Ex.~\ref{ex:ecomm1}, even if a session is established by the
dishonest broker, $\pmv A_1$ will not pay, and he will not be considered culpable 
for that. 
Indeed, the prefix 
$\ask{x}{\pmv A_1 \says \atom{pay}}$ is stuck because the contracts in
the session do not entail any obligation for $\pmv A_1$ to pay. 
For the same reason, $\pmv A_1$ will fulfil her contract
($\honest{C}{\pmv A_1}$).

In the following example, we show a different flavour of ``protection'':
the principals \pmv A and \pmv B are not protected by their contracts, 
but by the trusted escrow service that acts as a broker.

\begin{example} \label{ex:escrow-co2}
Recall the scenario in Ex.\ref{ex:escrow}.
The system is modelled as follows:

\[
\begin{array}{rcl}
  S & \mmdef & \pmv A[(x) (\tell {\pmv E} {\freeze {x} {( \atom{shipE}^+.\atom{pay}^-} )}.\
  \fact x {\atom{shipE}^+}.\
  \ask x {\Diamond{\atom{pay}^+}}.\
  \fact x {\atom{pay}^-})]
  \\
  & \mid &     \pmv B[(y) (\tell {\pmv E} {\freeze {y} {( \atom{payE}^+.\atom{ship}^-} )}.\
  \fact{y}{\atom{payE}^+}.\
  \ask{y}{\Diamond{\atom{ship}^+}}.\
  \fact x {\atom{ship}^-})]
  \\
  & \mid &     \pmv E[(z)(\ \tell{\pmv E}{\freeze {z} c_{\pmv E}}.\ \fuse z {\phi'}.\ P )]
  \\
  c_{\pmv E} & \mmdef & \atom{shipE}^-.\ \atom{payE}^-.\ ( \atom{pay}^+ \mid \atom{ship}^+) \quad + \quad
  \atom{payE}^-.\ \atom{shipE}^-.\ (\atom{pay}^+ \mid  \atom{ship}^+)
\end{array}\]
where $P$ is the obvious realization of the $c_{\pmv E}$, and 
$\phi' =  \Diamond(\atom{shipE}^+ \land  \Diamond\atom{pay}^+)
\land \Diamond( \atom{payE}^+ \land \Diamond  \atom{ship}^+)$.

The escrow service guarantees each of the
participants \pmv A and \pmv B that the other party has to fulfil
its obligation for the contract to be completed.
  The systems $S$ can perform the wanted interaction:
  \[
  S \to^* (s)(\ \pmv A[\nil] \mid \pmv B[\nil] \mid \pmv E[\nil] \mid s[\nil] \ )
  \]
\end{example}


\subsection{Variants to the basic calculus}\label{sec:variants}

Several variants and extensions are germane to \coco\!. 
We mention a few below.

\paragraph{Protection for contracts-as-processes.}
The contracts-as-processes model can be adapted so that \coco
processes can protect themselves from untrusted brokers.  In this
variant, contractual obligations would derive only from mutually {\em
compliant} contracts, i.e.~$c \vdash \phi$ should hold only when all
the (fair) traces of $c$ lead to $\nil$, in addition to $c
\models_{LTL} \phi$. Similarly, $\honest{c}{\pmv A}$ should also hold
on non-mutually-compliant $c$, since in this case no obligation
arises.
With this change, even if a principal $\pmv A$ is somehow put in a
session with fraudulent parties, $\pmv A$ can discover (via $\ask
{} {}$) that no actual obligation is present and avoid being swindled.

\paragraph{Contracts-as-processes with explicit sender and receiver.}
The syntax of contracts-as-processes (Def.~\ref{def:ccs-model}) can be
extended so to make explicit the intended senders/recipients of
inputs/outputs (permitting e.g.~to clearly state who is paying whom).
For this, we could use e.g.~$\atom{pay}^+@\pmv B$ as atoms, and adapt
the semantics $\xrightarrow{\mu}$ accordingly.

Note that the logic for observables $\Phi$ in Def.~\ref{def:ccs-model}
should be adapted as well.
Indeed, LTL does not distinguish between actions performed by distinct
principals.
To this purpose, we would allow $\pmv A \says \atom{a}^{+}@\pmv B$
(and related $\atom{a}^-,\atom{a}^0$ variants) as prime formulae, so
that it is now possible to observe who are the principals involved by
an action.

Note that the changes discussed above do not alter the general
calculus \coco\!, but merely propose a different contract model.

\paragraph{Local actions.}
In \coco\!, agents $\sys {\pmv A}{P}$ only carry latent contracts in
$P$.
Hence, communication between principals is limited to latent
contract exchange.
Allowing general data exchange in \coco can be done in a natural way
by following CCP.
Basically, $P$ would include CCP contraints $t$, ranging over a
constraint system $\tuple{T,\tilde\vdash}$, a further parameter to
\coco\!. Assume that $\pmv A \says t \in T$ for all $t\in T$ and for all $\pmv A$.
The following rules will augment the semantics of \coco 
(the syntax is extended accordingly):
{\small\[
\begin{array}{c}
      {\sys {\pmv A} {\tell {\pmv A} t. P + P' \mid Q} \xrightarrow{} 
       \sys {\pmv A} {\pmv A \says t \mid P \mid Q}}
    \\[10pt]
      {\sys {\pmv A} {\tell {\pmv B} t. P + P' \mid Q} \mid \sys {\pmv B} R \xrightarrow{} 
       \sys {\pmv A} {P \mid Q} \mid \sys {\pmv B} {\pmv A \says t \mid R}}
    \\[10pt]
    {\sys {\pmv A} {\ask {\pmv A} t . P + P' \mid T \mid Q} \redar \sys {\pmv A} {P \mid T \mid Q}}
    \mbox{\hspace{10pt} provided that \hspace{8pt}}
    {T \ \tilde\vdash\ t}

\end{array}
\]}
Note that the above semantics does not allow $\pmv A$ to corrupt the
constraint store of $\pmv B$ augmenting it with arbitrary constraints.
Indeed, exchanged data is automatically tagged on reception with the
name of the sender. So, in the worst case, a malicious $\pmv A$ can
only insert garbage $\pmv A \says t$ into the constraint store~of~$\pmv B$.

\paragraph{Retracting latent contracts.}
A $\mathsf{retract}$ primitive could allow a principal $\pmv A$ to
remove a latent contract of hers after its advertisement.
Therefore, $\pmv A$ could change her mind until her latent contract is
actually used to establish a session, where $\pmv A$ is bound to her
duties.

\paragraph{Consistency check.}
The usual $\mathsf{check}\ t$ primitive from CCP, which checks the
consistency of the constraint store with $t$, can also be added to
\coco\!.
When $\mathsf{check}\ t$ is executed by a principal $\sys {\pmv
  A}{P}$, $t$ is checked for consistency against the constraints in
$P$.
Note that checking the whole world would be unfeasible in a
distributed system.

\paragraph{Forwarding latent contracts.}
A $\mathsf{forward}$ primitive could allow a broker $\pmv A$ to
move a latent contract from her environment to that of another
broker $\pmv B$, without tagging it with $\pmv A \says$. 
In this way $\pmv A$ delegates to $\pmv B$ the actual opening of a session.

\paragraph{Remote queries.}
More primitives to access the remote principals could be added. Note
however that while it would be easy e.g.~to allow $\sys {\pmv A} {\ask
  {\pmv B} t}$ to query the constraint of $\pmv B$, this would
probably be undesirable for security reasons. Ideally, $\pmv B$ should
be allowed to express whether $\pmv A$ can access his own constraint
store. This requires some access control mechanism.


\section{Conclusions}

We have developed a formal model for reasoning about contract-oriented
distributed programs. 
The overall contribution of this paper is a
contract calculus (\coco) that is parametric in the choice of contract model. 
In \coco\!, principals can advertise their own contracts, find
other principals with compatible contracts, and establish a new
multi-party session with those which comply with the global
contract. 
We have set out two crucial issues: how to reach agreements, and how to detect
violations.
We have presented two concretisations of the abstract contract model. 
The first is an instance of the contracts-as-processes
paradigm, while the second is an instance of the
contracts-as-processes paradigm.

As a first step towards relating contract-as-processes and
contracts-as-formulae, we have devised a mapping from contracts based
on the logic PCL~\cite{BZ10lics} into CCS-like contracts. One
can then use contract-as-formulae at design-time, reason about them
using the entailment relation of PCL, and then concretise them to
contracts-as-processes through the given
mapping. Theorem~\ref{th:pcl-ccs} guarantees that
contracts-as-processes can reach success in those cases in which an
agreement would be possible in the logic model, hence providing a
connection between the two worlds.

\paragraph{Acknowledgments.}
This work has been partially supported by
Autonomous Region of Sardinia Project L.R.~7/2007 {\sc TESLA},
by PRIN Project {\sc Soft} (Tecniche Formali Orientate alla Sicurezza)
and by the Leverhulme Trust Programme Award ``Tracing Networks''.

\bibliographystyle{eptcs}
\bibliography{main}

\end{document}